\documentclass{ws-rv961x669}
\usepackage[square]{ws-rv-van}
\usepackage{ws-rv-thm}     
\usepackage{subfigure}     
\makeindex

\def\oq{\char'134}
\def\z0{\rm Z^0}
\def\epem{\rm e^+e^-}
\newcommand{\as}{\alpha_{\rm s}}

\begin{document}

\chapter[Look how it runs!]{Look how it runs!\label{ra_ch3}}

\author[S. Bethke]{Siegfried Bethke}

\address{Max-Planck-Institute of Physics, \\ 
F\"ohringer Ring 6, \\ 80805 Munich, Germany, \\
bethke@mpp.mpg.de
}
\begin{abstract}
Quantum Chromodynamics 
predicts that the strong
coupling strength $\as$ decreases with increasing energy 
or momentum transfer, and 
vanishes at asymptotically high energies.
The history and the status of experimental tests of
asymptotic freedom are summarised in this review.

\end{abstract}


\body


\section{Introduction}\label{ra_sec1}
Quantum chromodynamics (QCD) is the quantum field theory 
of the strong interaction between quarks mediated by gluons.
It is the strongest of the four fundamental forces of nature: 
between quarks and
at sub-atomic distances, the strong interaction is a hundred times stronger than
the electro-magnetic force, $10^{14}$ times stronger than the
weak force, and a mind-breaking
$10^{40}$ times stronger than gravity\footnote{A factor of $10^{40}$ 
corresponds to the difference in size between the
visible universe ($10^{26}$m) and an atomic nucleus ($10^{-14}$m).}.

QCD, exhibiting colour triplet quarks and colour octet gluons, was developed and introduced
by Harald Fritzsch, Murray Gell-Mann and Heinrich Leutwyler in 1972/1973 
\cite{fritzsch,fgl}.
One of the key features of non-abelian gauge theories (like QCD), 
{\em asymptotic freedom}, was discovered by David Gross, Frank Wilczek and David Politzer in 1973
\cite{qcd}.
Asymptotic freedom determines that the strong coupling strength, $\as$, \oq runs" with
energy and asymptotically vanishes at high energies.
This allows to calculate physical quantities by means of perturbation
theory, in regions of energies where $\as$ is smaller than 1.
After successful experimental verification of the running of $\as$, 
the discovery of Asymptotic Freedom was honoured with
the Nobel Prize in 2004 \cite{nobel2004}.

On the occasion of the {\em Symposium in Memory of Harald Fritzsch} held at the
Arnold-Sommerfeld Centre at Munich in July 2023.
a personal (re-)view of experimental tests of
asymptotic freedom in QCD is given,
recapping and updating a more extended article that was published in 2006 \cite{sb-2006}.

\section{Asymptotic freedom in QCD}

In QCD, the energy dependence of the running coupling $\as (\mu)$ is given by
the renormalisation group equation
\begin{equation} \label{eq-betafunction}
\beta (\as ) = \mu^2 \frac{d\as}{d\mu^2} =
- (b_0 \as^2 + b_1 \as^3 + b_2 \as^4 + ...) 
\end{equation}
with
\begin{equation}
b_0 = \frac{11 C_A - 2 N_f}{12 \pi}\ , \nonumber 
\end{equation}
where $C_A = 3$ is the number of \oq colour" charges in QCD, and $N_f$ is the
number of quark flavours produced at the energy scale $\mu$.
The higher order coefficients $b_1,\ b_2,\ b_3$ and $b_4$ are also known to date, see e.g.
\cite{Baikov2017}.

The \oq$-$" sign in front of the right-hand-side expression of eq.~\ref{eq-betafunction}
establishes asymptotic freedom, i.e. the running and decrease of $\as$ with increasing energy, as
long as $N_f  < 17$.

In 1-loop (leading order) approximation, eq.~\ref{eq-betafunction} is solved by

\begin{equation} \label{asq}
\as (\mu^2) = \frac{1}{b_0 \ln (\mu^2 / \Lambda^2)},
\end{equation}
\noindent
so QCD predicts that $\as$ logarithmically decreases with increasing energy scales $\mu$.

For a summary and discussion 
of higher order corrections, the role and treatment of quark production thresholds
and other peculiarities of QCD see e.g. \cite{sb-2006}.

\section{Experimental tests}

Since the late 1980's,
studies of hadron jets and determinations of $\as$, in wide ranges
of centre of mass energies and with increasing experimental
and theoretical precision, provided growing evidence for
asymptotic freedom.
Equally important and equivalent to the running coupling, evidence for
colour charged gluons and the SU(3) gauge structure as predicted by QCD 
provided key-indicators for the correctness of this theory describing the dynamics
of the strong interaction between quarks and gluons.

\begin{figure}
\centerline{\includegraphics[width=10.cm]{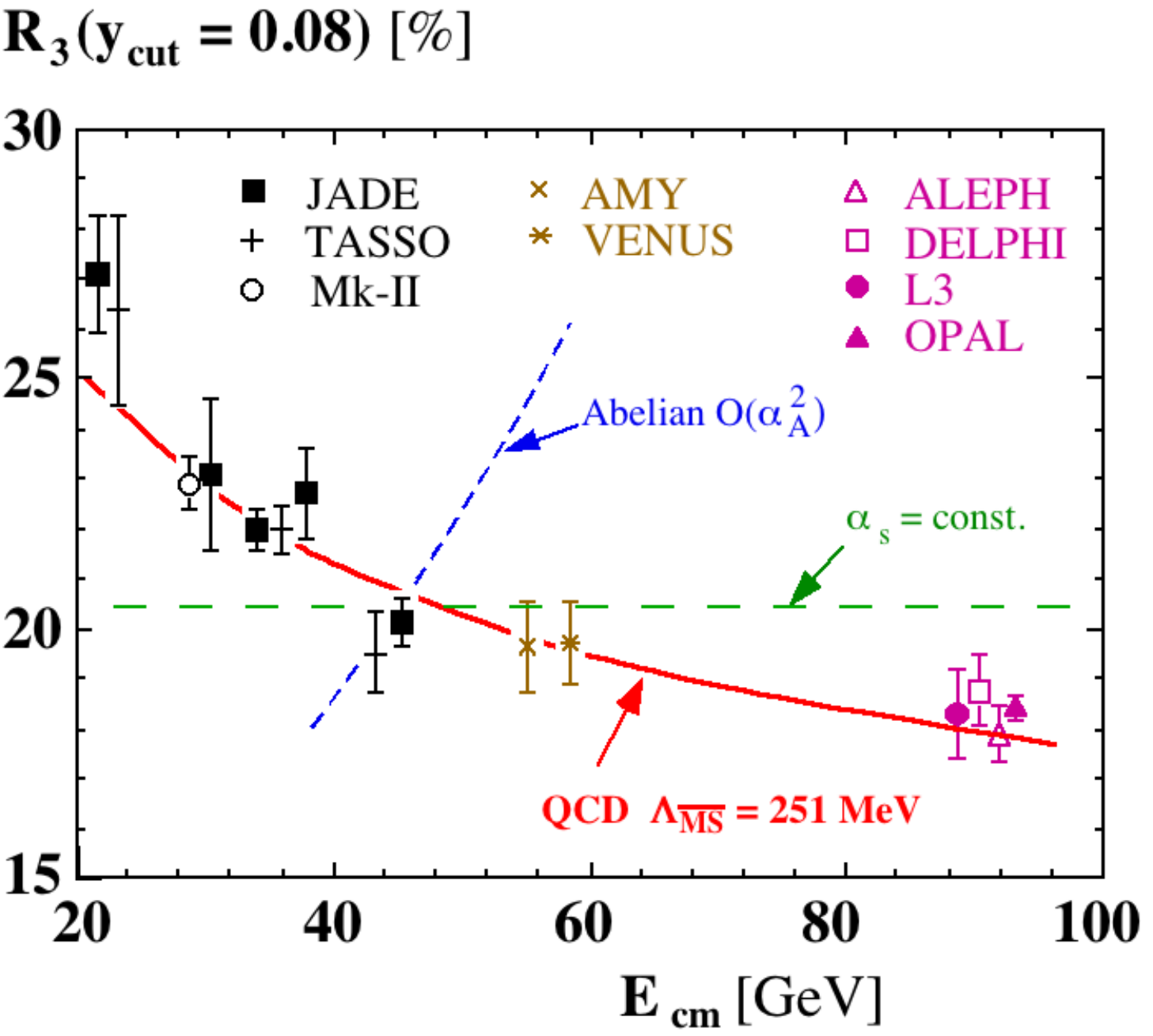}}
\caption{Energy dependence of 3-jet event production rates in
hadronic final states from $\epem$ annihilation,
measured using the JADE jet finder at a scaled jet
energy resolution $y_{cut} = 0.08$.
They are compared 
to theoretical expectations of QCD, of an abelian vector gluon model,
and to the hypothesis of a constant coupling strength.
\label{fig:R3-08}}
\end{figure}

\subsection{Energy dependence of jet production rates}

Studies of multi-jet event production rates in high-energy $\epem$ 
annihilation reactions, $\epem \rightarrow q \overline{q} (g) \rightarrow \rm{hadrons}$,
provided first evidence for the energy dependence of $\as$, before summaries
of explicit determinations of $\as$ could provide a convincing case for asymptotic freedom.
These studies used hadron jets as footprints of the initially produced
quarks and gluons.
They were based on the definition of resolvable hadron jets which could be applied
to both perturbative QCD calculation and to measured hadronic final states:

Within the JADE jet algorithm \cite{jadejet1}, the scaled invariant pair mass of two resolvable 
jets $i$ and $j$, $y_{ij} = M^2_{ij} / E^2_{vis}$ is required to exceed a threshold value $y_{cut}$, 
where $E_{vis}$ is the sum of all particles (resp. quarks and gluons) of the hadronic final state.
Theory predicts that - in leading order - the relative production rate of 3-jet events is
proportional to $\as$, 
\begin{equation}
R_3 = \frac{\sigma (\epem \rightarrow 3-jets)}{\sigma 
(\epem \rightarrow hadrons)} = C_1\left( y_{cut} \right) \as (\mu) , \nonumber
\end{equation}
\noindent 
plus higher order corrections $C_2(y_{cut}) \as^2(\mu) + \cal{O}$ $(\as^3)$,
where the coefficients $C_1$, $C_2$ etc. are constants, independent of the energy scale, and
can be calculated in QCD perturbation theory.
Model studies showed that corrections due to non-calculable hadronisation
effects and to limited detector resolution and acceptance are small.

The first study of the energy dependence of 3-jet
event production rates, at c.m. energies between 22 and 46 GeV,
analysed for constant jet resolution
$y_{cut}$ at the $\epem$ collider PETRA, 
provided first evidence for the energy dependence of $\as$
in 1988 \cite{jadejet2}.
These data are shown in figure~\ref{fig:R3-08}, together with 
more results from experiments at the PEP, TRISTAN \cite{sb-moriond-88}
and finally,
at the LEP collider \cite{sb-moriond-96}.
The measured 3-jet rates significantly decrease with 
increasing centre of mass energy, in excellent agreement
with the decrease predicted by QCD.
The hypothesis of an energy independent coupling, and especially
the prediction of an alternative, QED-like abelian vector gluon model,
where gluons carry no colour charge,
are in apparent contradiction with the data.

\begin{figure}[ht]
\centerline{\includegraphics[width=10.cm]{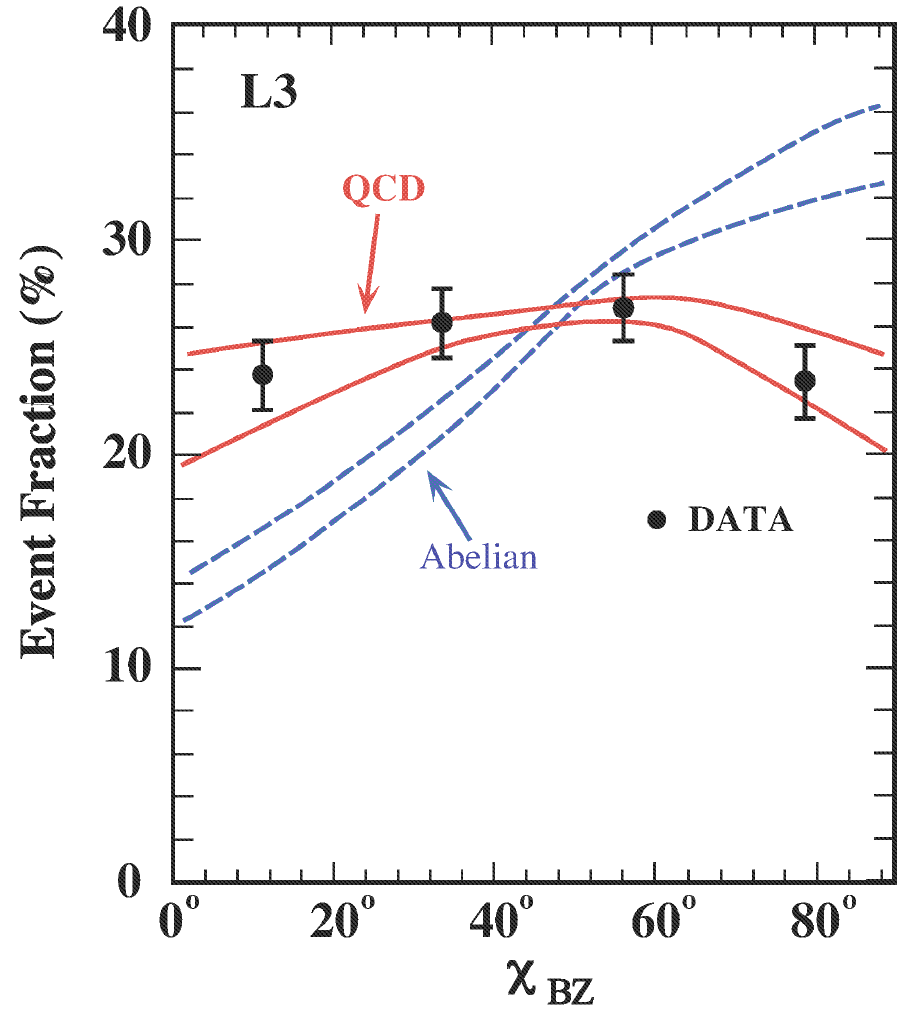}}
\caption{Distribution of the azimuthal angle between two planes
spanned by the two high- and the two low-energy jets of hadronic
4-jet events measured at LEP \cite{L3trip}, compared to the
predictions of QCD and of an abelian vector gluon model
where gluons carry no colour charge \cite{sbzerwas}.
\label{fig:L3trip}}
\end{figure}

\subsection{Gluon self-coupling and QCD group structure constants}

The gluon self-coupling is a direct consequence of gluons carrying
colour charge by themselves, and is the essential feature leading to the prediction of
asymptotic freedom. 
Methods to directly detect effects of gluon self-coupling
are based on analysing distributions
which are sensitive to the spin structure of hadronic 4-jet final states, see e.g.
\cite{sbzerwas}. 
The Bengtson-Zerwas angle, $\chi_{BZ}$
\cite{bz},
measuring the angle between the planes defined by the two
highest and the two lowest energy jets, is rather sensitive to
the difference of a gluon splitting into two gluons, 
which in QCD is the dominant source of 4-jet final states, and a gluon splitting
to a quark-antiquark pair, which is the dominant process in an
abelian vector theory where gluons carry no colour charge.

The results of an early study which showed convincing evidence for
the gluon self coupling in 1990 \cite{L3trip}, after one year
of data taking at the $\epem$ collider LEP, 
is shown in figure~\ref{fig:L3trip}.
The data clearly favour the QCD prediction and rule out the abelian
vector gluon case.

\begin{figure}[ht]
\centerline{\includegraphics[width=10.cm]{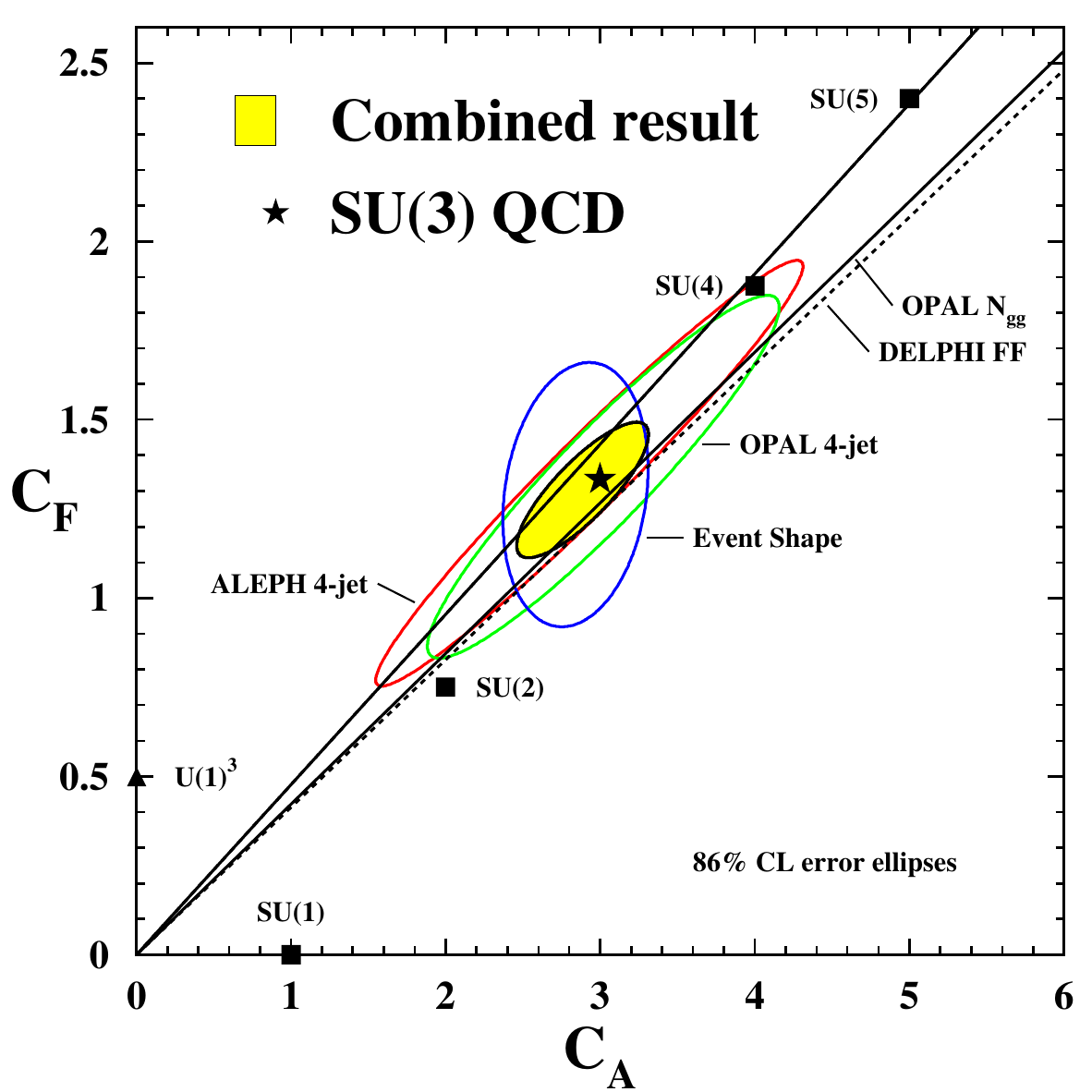}}
\caption{Determinations of the QCD colour factors $C_A$ and $C_F$ \cite{kluth-tgv}.
\label{fig:cacfplot}}
\end{figure}

The QCD group constants
$C_A$, $C_F$ and $N_f$, assume values of 3, 4/3 and 5 in
a theory exhibiting SU(3) symmetry, with 5 quark flavours
in the vacuum polarisation loops.
For alternative, QED-like toy models of the strong
interaction with U(1) symmetry, these values would be 
$C_A=0$ and $C_F=0.5$.
Experimental determination of these two parameters thus constitute
one of the most intimate tests of the predictions of QCD.

At LEP, data statistics and precision allowed to actually
determine experimental values for $C_A$, which basically is the 
number of colour charges, and $C_F$.
A summary of such studies, involving  
analyses of 4-jet angular correlations 
and fits to hadronic event shapes \cite{kluth-tgv}, is presented  
in figure~\ref{fig:cacfplot}.
The data are in excellent agreement with the gauge structure constants of 
QCD. 
They rule out the Abelian vector 
gluon model and theories exhibiting symmetries other than SU(3).

\subsection{Energy dependence of $\as$}
\label{determinations}

The {\it straight forward way}, clearly, is to determine $\as (Q)$
from data at different and large ranges of energy scales $Q$.
$\as$, however, is not an observable by itself, but must be determined 
through comparison of measurements of hadronic final states of high energy reactions
with QCD predictions of the dynamics of quarks and gluons. 
Such determinations became 
available in the early 1980s,
but their significance was rather restricted due to limited statistical precision,
limited precision of theoretical calculations in only leading or next-to-leading
order perturbation theory, and the uncertainties in
parametrising the hadronisation process, i.e.
the transition from quarks and gluons to observable hadrons.

\begin{figure}
\centerline{\includegraphics[width=10.cm]{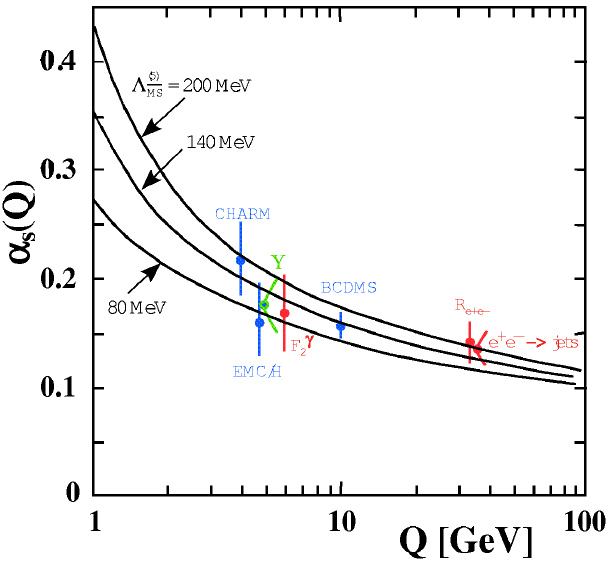}}
\caption{Summary  of $\as$ determinations 
in the year 1989 \cite{altarelli89}, with data
from deep inelastic lepton-nucleon scattering (blue),
heavy quarkonia (green)
and $\epem$ annihilation experiments (red).}
\label{fig-as89}
\end{figure}

The status of early determinations of $\as$, as summarised in 1989 \cite{altarelli89},
is reproduced in Fig.~\ref{fig-as89}.
While being compatible with the QCD expectation, the limited
energy range, data statistics and theory precision, at that time, were
not yet sufficient to claim experimental evidence for the running of $\as$.

The situation rapidly improved with the start-up of the LEP $\epem$ collider
at CERN in 1989.
By 1992, LEP provided significant results on $\as$ at energies around the 
mass of the $Z^0$ resonance, $M_Z \sim 91.2$~GeV/$c^2$, from hadronic event shapes 
from the decay width of the $Z^0$-boson, and from hadronic decays of the $\tau$-lepton,
at the energy scale of $M_{\tau} \sim 1.78$~GeV/$c^2$ - thus expanding the available energy range
by almost an order of magnitude.
Precision and reliability of theoretical calculations also improved by that time,
in terms of next-to-next-to-leading order (NNLO) precision for hadronic $Z^0$ and $\tau$
decays, and resummation of NLO predictions for hadronic event shapes and jet production rates.
These ingredients gradually led to emerging evidence for the running of $\as$, in accordance with QCD
and Asymptotic Freedom, see e.g. \cite{sb-catani} for a summary of $\as$ results at that time.

\begin{figure}[ht]
\centerline{
  \minifigure[Summary of $\as$ determinations in 2002 \cite{as2002}]
     {\includegraphics[width=6.1cm]{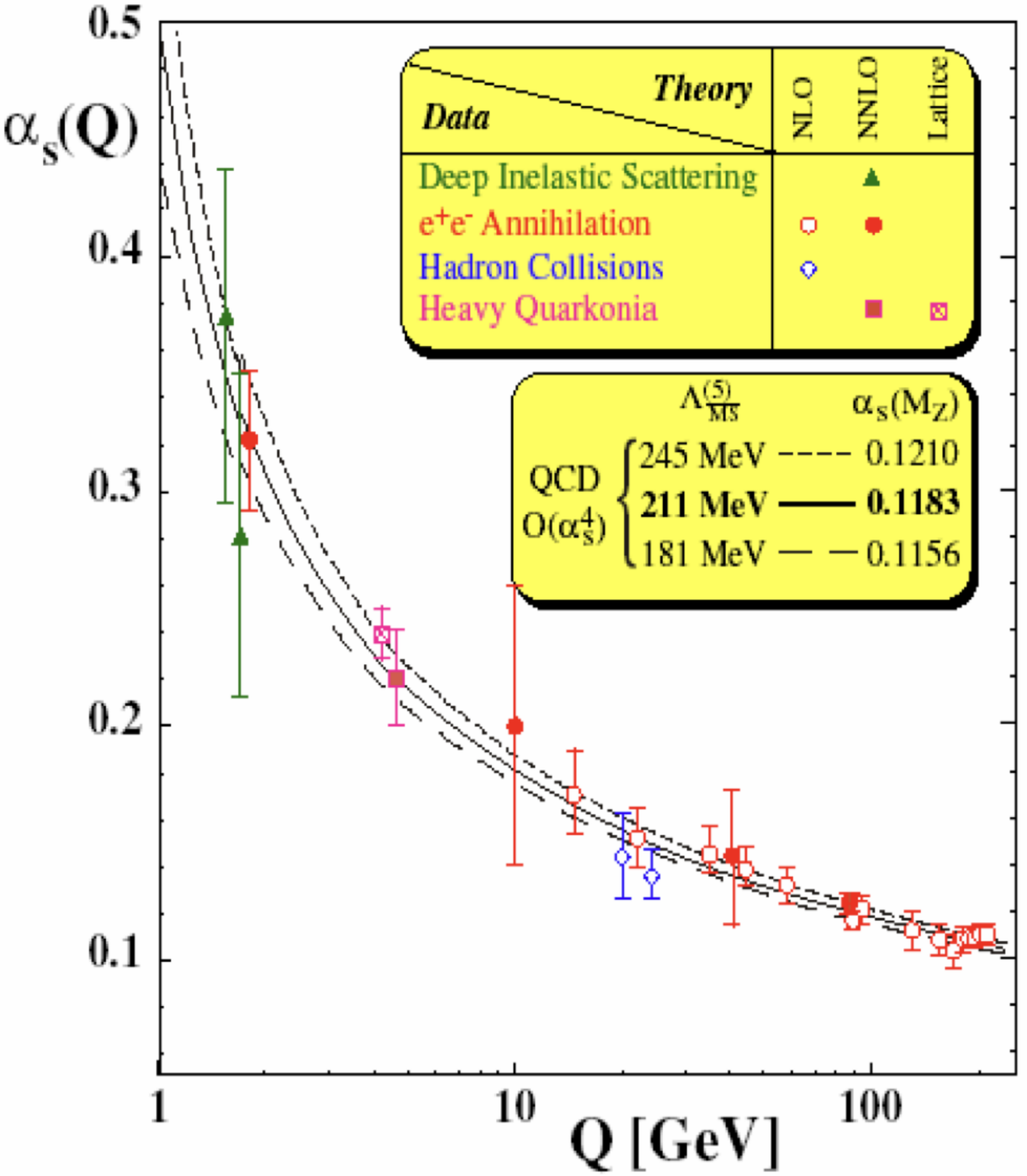}\label{as2002}}
  \hspace*{4pt}
  \minifigure[The running $\as$ from low energy PETRA to high energy LEP data \cite{aszerwas}.]
     {\includegraphics[width=5.2cm]{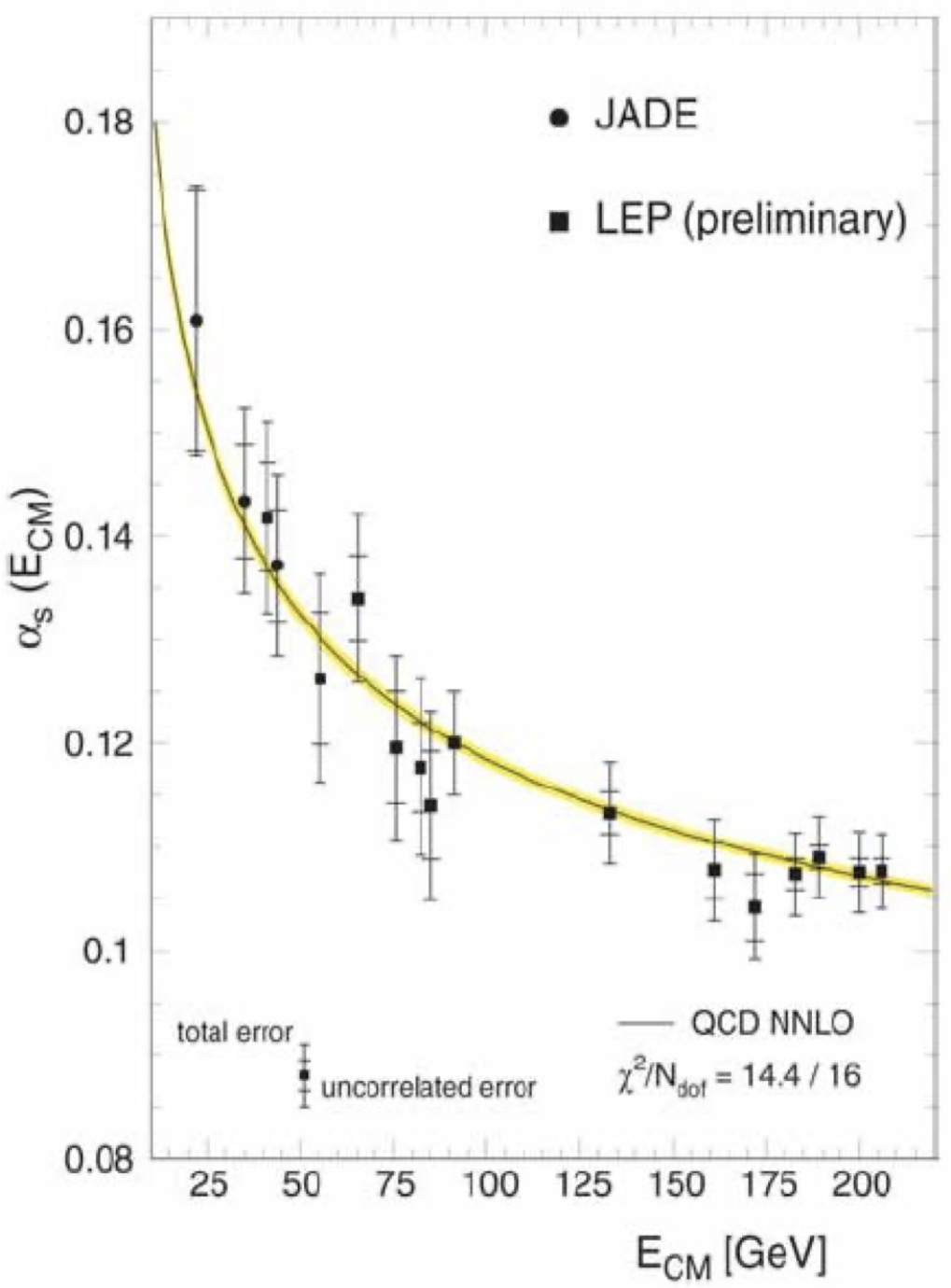}\label{aszerwas}}
}
\end{figure}

Ten further years later, by 2002 - about 30 years after the birth of QCD~-, 
increased precision of results from LEP,
re-application of up-to-date analysis methods and theoretical developments on past data
from the JADE experiment at PETRA, 
and precise results from lattice QCD calculations 
together provided convincing evidence for the running of $\as$, over 
a range of energy scales
from 1.7 to 214~GeV, in perfect agreement with the expectation from QCD,
see figs. \ref{as2002} and \ref{aszerwas}.
The evidence, at that point, was strong  enough to justify the award of the
2004 Nobel Prize in Physics to David Gross, Frank Wilczek and David
Politzer \oq for the discovery of asymptotic freedom in the theory of the strong
interaction" \cite{nobel2004}.
In fact, in the official advanced information on the scientific background of
this Prize \cite{nobel2004b}, figs. \ref{as2002} and \ref{aszerwas} are presented 
as $the$ experimental evidence for asymptotic freedom in QCD.

\begin{figure}
\centerline{\includegraphics[width=11.cm]{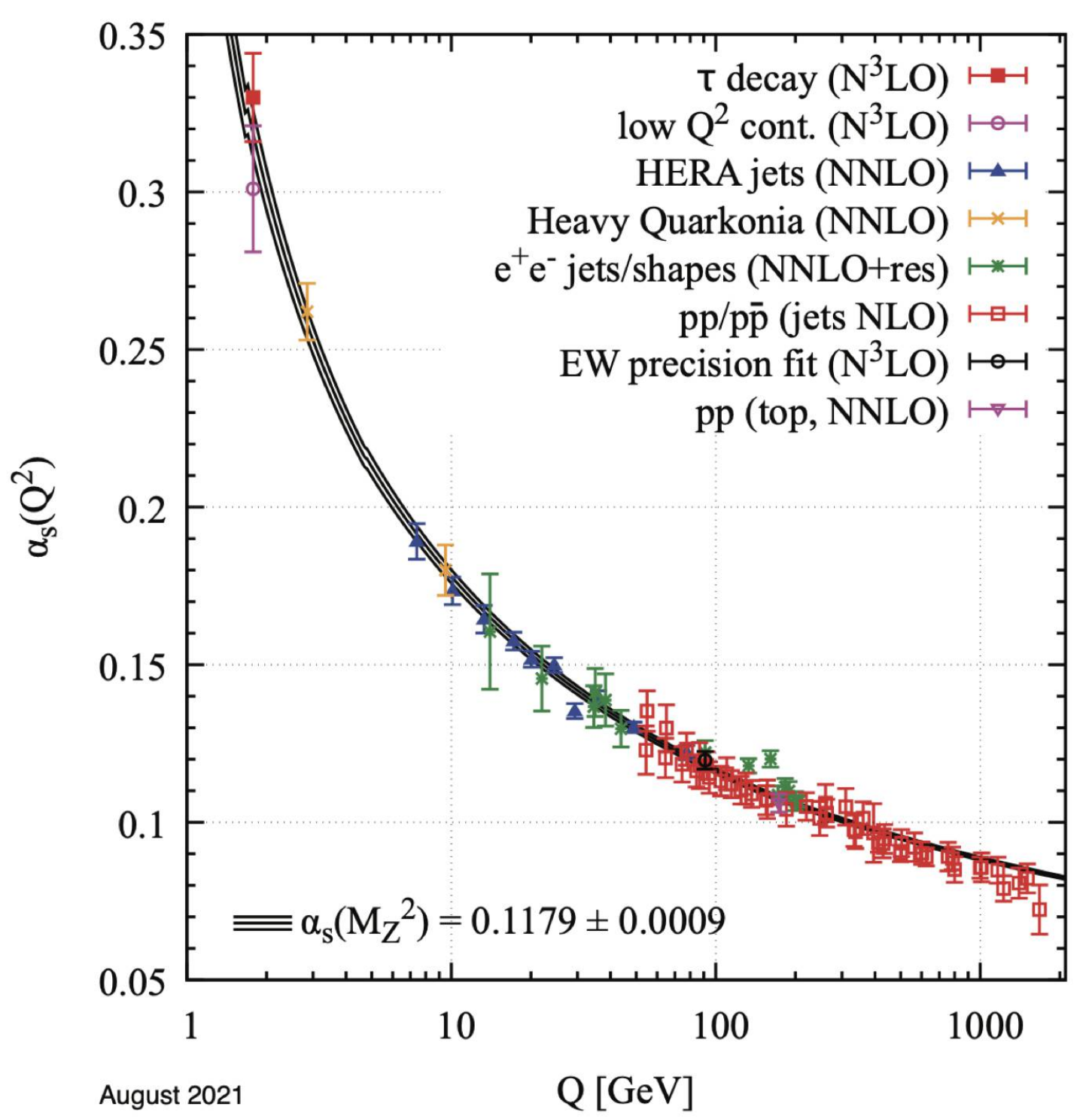}}
\caption{Summary  of $\as$ determinations 
in the year 2021 \cite{rpp},
compared with the QCD prediction of the running coupling,
in 4-loop precision plus 3-loop matching at the charm- and bottom-quark 
thresholds. }
\label{as2021}
\end{figure}

The 2004 Nobel Prize was not the end of demonstrating asymptotic freedom in QCD:
many new and significant results were obtained since then - now about 50 years after the birth
of QCD, and more than 30 years after Altarelli's review (c.f. fig.~\ref{fig-as89}).
Precise determinations of $\as$ from lepton-hadron collisions at HERA and from hadron
collisions at the Tevatron and the Large Hadron Collider (LHC) became available,
now covering the huge energy range from 1.7 to more than 2000~GeV.
In many cases, experimental uncertainties are at and below the one per-cent level.
Theory predictions are up to the level of N$^3$LO, and lattice calculations 
also reached precisions of the order of one per-cent and below.
An up-to-date summary of all results \cite{rpp}
is reproduced in fig.~\ref{as2021}.
It provides a very precise confirmation of the QCD prediction of
asymptotic freedom and the specific functional form of $\as$
logarithmically decreasing with increasing energy scale, c.f.
equation \ref{asq}.
No other reasonable parametrisation can provide a similarly accurate description
of the data.

\section{Conclusion}

Fifty years after the foundation of quantum chromodynamics,
the experimental evidence for the key-feature of QCD, asymptotic
freedom and the logarithmic decrease of the coupling $\as$ with
energy, is overwhelming and leaves no doubt on QCD being the
gauge theory correctly describing the strong interaction between quarks
and gluons.
As a well-tested and \oq proven" theory, its predictive power now serves as
an indispensable and reliable source for defining and refining 
the \oq standard model" hadronic contributions and background to many
other precision studies, like electro-weak phenomena and searches 
for new physics.

\section*{Acknowledgements}
Harald Fritzsch was one of the founding pioneers of QCD.
Thank you Harald, for your colossal contributions and lasting impact
to the field!

\bibliographystyle{ws-rv-van}
\bibliography{ws-rv-sample}

\end{document}